\documentclass[prd,twocolumn,floats,floatfix,nofootinbib]{revtex4}
\usepackage[dvips]{graphicx}
\usepackage{graphics}
\usepackage{dcolumn}
\usepackage{amssymb}
\usepackage{bm}
\usepackage{amsmath}
\usepackage{color}
\usepackage{epstopdf}
\usepackage{multirow}
\usepackage{ulem}

\def\spose#1{\hbox to 0pt{#1\hss}}

\def\lta{\mathrel{\spose{\lower 3pt\hbox{$\mathchar"218$}}
     \raise 2.0pt\hbox{$\mathchar"13C$}}}
\def\gta{\mathrel{\spose{\lower 3pt\hbox{$\mathchar"218$}}
     \raise 2.0pt\hbox{$\mathchar"13E$}}}
\newcommand{\be}{\begin{equation}}
\newcommand{\en}{\end{equation}}
\newcommand{\bea}{\begin{eqnarray}}
\newcommand{\ena}{\end{eqnarray}}

\usepackage[dvipsnames]{xcolor}    

\def\revised#1{\textcolor{black}{#1}}

\begin{document}
 
\title{Nonsingular Cosmology from an Interacting Vacuum}

\author{Marco Bruni${{^{a,b}}}$\footnote{marco.bruni@port.ac.uk}, Rodrigo Maier${{^c}}$\footnote{rodrigo.maier@uerj.br}, David Wands${{^a}}$\footnote{david.wands@port.ac.uk}
\vspace{0.5cm}}

\affiliation{${{^a}}$Institute of Cosmology {\rm \&} Gravitation, University of Portsmouth, Dennis Sciama Building, Burnaby Road, Portsmouth, PO1 3FX, United Kingdom\\
\\
${{^b}}$INFN Sezione di Trieste, Via Valerio 2, 34127 Trieste, Italy\\
\\
${{^c}}$Departamento de F\'isica Te\'orica, Instituto de F\'isica, Universidade do Estado do Rio de Janeiro,\\
Rua S\~ao Francisco Xavier 524, Maracan\~a,\\
CEP20550-900, Rio de Janeiro, Brazil\\
}


\date{Received 11 November 2021; accepted 10 March 2022; published 28 March 2022}

\begin{abstract}
We examine the dynamics of FLRW cosmologies in which the  
vacuum interacts with a perfect fluid through an energy exchange, \revised{focusing on the exploration of} nonsingular configurations, including cyclic and bouncing models.
We consider two specific choices for the energy transfer. 
In the first case, the energy transfer is proportional to
a linear combination of the vacuum and fluid energy densities
which makes the conservation equations exactly integrable. 
The resulting Friedmann equation 
can be interpreted as an energy constraint equation with an effective potential for the scale factor that may include an infinite barrier forcing a bounce at small values of the scale factor, as well as a potential well allowing for cycling solutions.
In the second case, the energy transfer is a nonlinear
combination of the vacuum and fluid energy densities. Nonetheless even in this case 
the dynamics can be partially integrated, leading to a first integral,  reducing the number of degrees of freedom. We show that also in this nonlinear case bouncing and cycling cosmologies may arise. 
In both cases the structure of the resulting phase space allows for nonsingular orbits with an early accelerated phase around a single bounce, connected via a decelerated matter-dominated era to a late-time accelerated phase dominated by an effective cosmological constant.

\end{abstract}
\maketitle
\section{Introduction}

Although General Relativity is the most successful theory that currently describes gravitation, it is well understood
that it generally gives rise to singular solutions at high energies.
Indeed, despite the successes of the standard $\Lambda$CDM model in describing the evolution of the Universe
and its current state\cite{weinbergtb,maartenstb,Jones:2017xsc}, the initial singularity still presents an obstacle to properly understanding gravitation at the highest energy scales. 
During recent decades different theories of gravitation have been considered in order to solve the initial singularity 
problem by modifying General Relativity in the deep ultraviolet regime. In this context, bouncing
models have been proposed to circumvent the flatness/horizon problems of the standard hot big bang cosmology and reproduce the power spectrum of primordial cosmological perturbations inferred by observations~\cite{Gasperini:1992em,Khoury:2001wf,Kallosh:2001ai,Steinhardt:2001st,Wands:2008tv,Maier:2013hr,Maier:2011yy}.

On the other hand, over the past twenty years or more observational data~\cite{SupernovaSearchTeam:1998fmf, SupernovaCosmologyProject:1998vns, Rubin:2008wq, Hicken:2009dk, BOSS:2016wmc, Planck:2018vyg} have given support to the idea that our Universe is currently in a state of accelerated expansion. In order to explain such behaviour, a new field -- known as dark energy -- 
that violates the strong energy condition \cite{Hawking:1973uf,Wald:1984rg,maartenstb,Visser:1997qk} in the deep infrared,  i.e.\ in the late universe, has been considered.
Although the cosmological constant
seems to be the simplest and most appealing candidate for dark energy, it poses a severe problem to quantum field
theory to accommodate its observed tiny value with theoretical calculations of its vacuum energy~\cite{Weinberg:1988cp}.
Different candidates for dark energy have also been proposed in the realm of modified theories of gravitation \cite{amendola}.

The appearance of a cosmological singularity in General Relativity is typically due to assuming standard energy conditions \cite{Hawking:1973uf,Wald:1984rg,maartenstb} that can be violated by dark energy \cite{Visser:1997qk}. Thus it is worth reconsidering the high-energy regime in General Relativity, to see if the singularity can be avoided by some form of dark energy dominating in this high-energy regime~\cite{Carneiro:2009et,Chimento:2014tua}, possibly producing a bounce \cite{Ananda:2005xp, Ananda:2006gf,Ganguly:2019llh}. 

Extending the above scenario, the possibility of an interacting component, with vacuum equation of state, $w=-1$, has been a subject of considerable interest \cite{Bertolami:1986bg,Alcaniz:2012mh,Freese:1986dd,Carvalho:1991ut,Shapiro:2000dz,Wands:2012vg}
\revised{some of which has be motivated by quantum field theory considerations~\cite{Chen:1990jw,Lima:2013dmf,Moreno-Pulido:2020anb}}.
In the context of black hole formation it has been shown~\cite{Maier:2020bgm} that the collapse of barotropic perfect
fluids, namely dust and radiation, may give rise to Reissner-Nordstr\"om-de Sitter black holes
for an appropriate choice of the energy exchange between the nonrelativistic perfect fluid and the vacuum component.
From the cosmological point of view on the other hand, it has been shown that an
interacting dark energy component may also ease cosmological tensions between different observational datasets 
\cite{Wang:2013qy, Salvatelli:2014zta, Benetti:2021div, Wang:2015wga, Zhao:2017cud, Sola:2016sbt, DiValentino:2017iww, Kumar:2017dnp, Martinelli:2019dau, Hogg:2020rdp,Hogg:2021yiz,DiValentino:2021izs,SolaPeracaula:2021gxi}.

In this paper we address the issue of an interacting vacuum component in the framework of nonsingular 
cosmology. In section~\ref{sec:model} we present our interacting vacuum equations in which
we consider two distinct 
\revised{phenomenological} 
models  -- linear and nonlinear -- for the energy transfer between the vacuum component
and a barotropic fluid such as nonrelativistic matter or radiation. In section~\ref{sec:linear} we examine the linear case in which the full dynamics can be integrated and a modified Friedmann evolution is obtained including 
a correction term that leads to nonsingular solutions, some with  a single bounce in the early evolution
of the universe, some perpetually cycling between a bounce and a turn-around. Section IV is devoted to the case of a nonlinear interaction.
In this case we obtain a first integral of the dynamics, reducing the number of degrees of freedom.
The eigenvalues of the linearization matrix about fixed points in the phase space are evaluated in order to explore the existence of nonsingular configurations. We show that also in this nonlinear case bouncing and cycling models do exist. We summarise and present our conclusions in section V. We assume General Relativity and natural units where $c=1$.

\section{The interacting vacuum equations}
\label{sec:model}
We start by considering the Einstein
field equations
\begin{eqnarray}
\label{eqm1}
G_{\mu\nu}= \kappa^2 (T_{\mu\nu}-V g_{\mu\nu}),
\end{eqnarray}
where $G_{\mu\nu}$ is the Einstein tensor and
$\kappa^2=8\pi G$ is the Einstein constant.  $T_{\mu\nu}$ is the energy-momentum tensor for matter, which we will take to be a perfect fluid:
\begin{eqnarray}
\label{eqm3}
T^{\mu\nu}=(\rho+p)u^\mu u^\nu+ p g^{\mu\nu},
\end{eqnarray}
where $u^\mu$ is the $4$-velocity of the fluid,  $\rho=T_{\mu\nu}u^\mu u^\nu$ its rest-frame energy density and $p$ its pressure. The energy-momentum of the vacuum is also that of a perfect fluid with $p=-\rho$; denoting its energy density by $V$, this gives the $-Vg_{\mu\nu}$ term in \eqref{eqm1}. It follows that any 4-vector is an eigenvector for the vacuum energy-momentum tensor, with $V$ its energy density in the frame of any observer. The matter-vacuum
interaction 
is described by an energy-momentum transfer $4$-vector $Q_\nu$, so that the conservation equations for the two components are
\begin{eqnarray}
\label{eqm41}
\nabla_\mu (T^{\mu}_{~~\nu})&=&-Q_\nu,\\
\label{eqm42}
-\nabla_\nu V &=& Q_\nu,
\end{eqnarray}
where the equal and opposite signs for $Q_\nu$ are required by the Bianchi identities.

The $4$-vector $Q_\nu$ can in general be decomposed 
in two  parts, parallel and orthogonal to the $4$-velocity of the fluid,
\begin{eqnarray}
\label{eqq}
Q^\mu= Q u^\mu + q^\mu.
\end{eqnarray}
In the above $Q$ denotes an energy flow in the rest frame of the fluid, while $q^\mu$ is connected to momentum exchange
between matter and vacuum. 
In this paper we shall consider
the case in which the interaction reduces to a pure energy exchange~\cite{Salvatelli:2014zta, Wang:2015wga,Martinelli:2019dau,Hogg:2020rdp,Hogg:2021yiz}
so that $q^\mu =0$, simply because we shall focus on homogeneous-isotropic models where this restriction follows from symmetries. In this case, $Q^\nu$ is parallel to the matter $4$-velocity, $Q^\nu =Q u^\nu$,
and matter is not accelerated due to its interaction with the vacuum. In fact, if one assumes a non-relativistic perfect fluid, it can be shown that for $q^\mu =0$ the matter distribution remains geodesic~\cite{Wang:2013qy}. Constraints on the interacting vacuum in this geodesic CDM scenario were examined in~\cite{Salvatelli:2014zta, Wang:2015wga,Martinelli:2019dau,Hogg:2020rdp,Hogg:2021yiz}.

In this paper we will examine two different covariant choices for $Q$:
\begin{eqnarray}
\label{eqm111}
 Q_1=[\xi (V_\Lambda-V)+ \sigma \rho]\Theta,\\
\label{eqm112}
Q_2=\chi(1-V/V_{\Lambda})\rho \Theta.~~~~
\end{eqnarray}
In the above, $\Theta=\nabla_\mu u^\mu$ is the expansion scalar, and  $\xi$, $\sigma$ and $\chi$ are dimensionless
coupling parameters. 
%
In both cases $V_\Lambda$ plays the role of an effective cosmological constant\footnote{By this we mean that we don't have a $\Lambda$ term in Einstein equations, rather a cosmological constant appears as a fixed point of the vacuum dynamics.}, i.e., an asymptotic value of $V\to V_\Lambda$.

We will study the dynamics in a Friedmann-Lemaitre-Robertson-Walker (FLRW) spacetime where,
$\Theta=3H$, and $H\equiv \dot{a}/a$ is the Hubble rate. 
Choosing the equation of state $p=w \rho$ where $w$ is constant, equations (\ref{eqm41}) and (\ref{eqm42}) reduce to
\begin{eqnarray}
\label{eqm71}
\dot{\rho}+3(1+w)H\rho&=&-Q,\\
\label{eqm72}
\dot{V}&=&Q.
\end{eqnarray}
From the Einstein field equations on the other hand, we obtain
\begin{eqnarray}
\label{eqm8}
\dot{H}=-\frac{k}{2a^2}-\frac{3H^2}{2}+\frac{\kappa^2}{2}(V- w\rho).
\end{eqnarray}
Assuming that $Q\equiv Q(\rho, V, H)$, we see that (\ref{eqm71})-(\ref{eqm8}) constitute a nonautonomous dynamical system 
whose first integral is given by the Friedmann equation
\begin{eqnarray}
\label{eqm9}
H^2+\frac{k}{a^2}=\frac{\kappa^2}{3}(\rho +V).
\end{eqnarray}
This dynamical system can be turned into 
an autonomous configuration by substituting (\ref{eqm9}) in (\ref{eqm8}).
In this case, (\ref{eqm8}) can be rewritten as
\begin{eqnarray}
\label{eqm10}
\dot{H} =-H^2 -\frac{\kappa^2}{6}[\rho(1+3w)-2V],
\end{eqnarray}
which is the standard form of the Raychaudhuri equation in the case of a FLRW spacetime.

\revised{
Note that equation \eqref{eqm10} is even in $H$ (it depends only on $H^2$), while the energy transfer $Q$ in Eqs.~\eqref{eqm111} and~\eqref{eqm112} is proportional to $H$, so that in both cases, linear and nonlinear, the resulting coupled energy conservation equations~\eqref{eqm71} and~\eqref{eqm72} are also proportional to $H$. The net result of this is that the evolution during contraction ($H<0$) is the mirror image that during expansion ($H>0$). 
Thanks to the proportionality of the energy conservation equations to $H$, in both cases the $H$ dependence can be eliminated from the coupled equations, a fact that we are going to exploit in the following sections, and that implies an overall adiabatic evolution. 
As will be clear from the phase-space plots, the evolution of the homogeneous and isotropic models is completely reversible, with no entropy production and no arrow of time\footnote{\revised{This symmetry is typically broken by the evolution of inhomogeneities, even at first order in perturbations. In a FLRW background it can only be broken by a bulk viscosity contribution to the equation of state, phenomenologically represented by  $p_\theta=-\xi\Theta=-3\xi H$ ($\xi$ being a bulk viscosity parameter), or in the case of a scalar field, because the d'Alembertian operator in FLRW also contains a friction term proportional to $H$.} }.  }

\section{The Linear case \boldmath$Q=Q_1$}
\label{sec:linear}

In the case (\ref{eqm111}) of a linear interaction, equations (\ref{eqm71}) and (\ref{eqm72}) can be rewritten as
\begin{eqnarray}
\label{eqa11}
\dot{\rho}&=&-3H[(1+w+\sigma)\rho+\xi(V_\Lambda-V)],\\
\label{eqa12}
\dot{V}&=&{3H}[\sigma \rho+\xi(V_\Lambda-V)].
\end{eqnarray}
In the limit $V_\Lambda\to0$, and setting $w=k=0$, this reduces to the two-fluid cosmology studied in Ref.~\cite{Kaeonikhom:2020fqs}, where in that paper the linear interaction parameters were $\alpha\equiv3\sigma$ and $\beta\equiv-3\xi$. 
The system above can also be seen as a special case of the most general linear coupling of two cosmological fluids considered in \cite{Quercellini:2008vh}. \revised{The focus of \cite{Kaeonikhom:2020fqs,Quercellini:2008vh} was on studying these interaction models as a possible alternative to a cosmological constant as the simplest form of dark energy in the late universe, while our focus here is on the possible non-singular behaviour of these models at high energies. }

To integrate the full dynamics, we note that (\ref{eqa11}) and (\ref{eqa12})
correspond to a coupled system of linear first order ordinary differential equations for the functions
$\rho$ and $V$. By decoupling this system it can then be shown that the general solution is
\begin{eqnarray}
\label{eqa81n}
\rho &=& E_1 a^{\alpha_1}+E_2 a^{\alpha_2},\\
\label{eqa82n}
V&=&V_\Lambda+\lambda_1 a^{\alpha_1}+\lambda_2 a^{\alpha_2},
\end{eqnarray}
where
\begin{eqnarray}
\label{eqa91}
\alpha_1&=&-\frac{3}{2}(1+w+\xi + \sigma+\Delta),\\
\label{eqa92}
\alpha_2&=&-\frac{3}{2}(1+w+\xi + \sigma-\Delta),
\end{eqnarray}
are the roots of the characteristic polynomial for \eqref{eqa11} and \eqref{eqa12}
and
\begin{eqnarray}
\label{delta}
\Delta=\sqrt{(1+w+\sigma-\xi)^2+4\sigma\xi}.
\end{eqnarray}
Furthermore, the constants of integration $\lambda_1$, $\lambda_2$, $E_1$ and $E_2$ are subject to the constraints 
\begin{eqnarray}
\label{eqa101}
 \lambda_1 \xi - E_1 \Big(1+w+\sigma+\frac{\alpha_1}{3}\Big)=0,\\
\label{eqa102}
\lambda_2 \xi - E_2 \Big(1+w+\sigma+\frac{\alpha_2}{3}\Big)=0,
\end{eqnarray}
or, equivalently,
\begin{eqnarray}
\label{vinc1}
\lambda_1\Big(\frac{\alpha_1}{3}+\xi\Big)-\sigma E_1=0,\\
\label{vinc2}
\lambda_2\Big(\frac{\alpha_2}{3}+\xi\Big)-\sigma E_2=0.
\end{eqnarray}
From (\ref{delta}) we see that the necessary and  sufficient condition for the existence of two linearly independent and  real solutions in (\ref{eqa81n}) and (\ref{eqa82n}), i.e.\ to have $\alpha_1\not =\alpha_2$ real and distinct,
is given by
\begin{eqnarray}
\label{pds1}
[\xi-(1+w+\sigma)]^2+4\sigma\xi > 0.
\end{eqnarray}
Note that we are explicitly excluding the case $\Delta=0$, for in this case we would not have the two independent solutions given by $\alpha_1\not = \alpha_2$. It is also worth pointing out
that the solutions (\ref{eqa81n}) and (\ref{eqa82n}) are obtained directly from integrating the continuity equations (\ref{eqa11}) and (\ref{eqa12}) and do not use the Friedmann constraint. Hence they are valid for homogeneous cosmologies of arbitrary $3$-curvature, including anisotropic cosmologies where $a^3=\int\Theta dt$ is the generalised volume factor, in any theory of gravity.

In order to illustrate the behaviour
of $\alpha_1$ and $\alpha_2$ as functions of $\xi$ and $\sigma$, let us consider the simple case in which $w=0$.
In order to reproduce in Eq.~(\ref{eqa81n}) the evolution of the matter density found in a standard non-interacting matter-dominated epoch, $\rho\propto a^{-3}$ for some period, we require that
either $\alpha_1=-3$ or $\alpha_2=-3$.
In the case $\alpha_1=-3$, we obtain $\sigma=0$ and $\Delta=1-\xi$, and hence we have $\alpha_2=-3\xi$. 
On the other hand, fixing $\alpha_2=-3$ we obtain                                          $\sigma=0$ and $\Delta=\xi-1$, and hence $\alpha_1=-3\xi$.

\begin{figure}
\includegraphics[width=6cm,height=7cm]{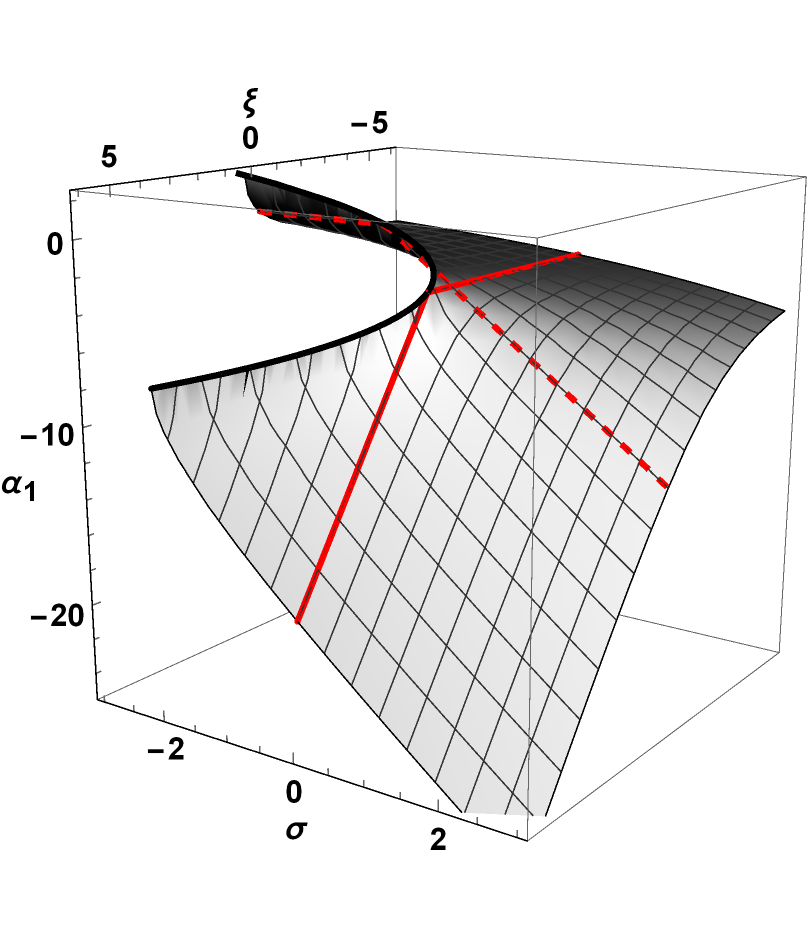}
\includegraphics[width=5.8cm,height=6cm]{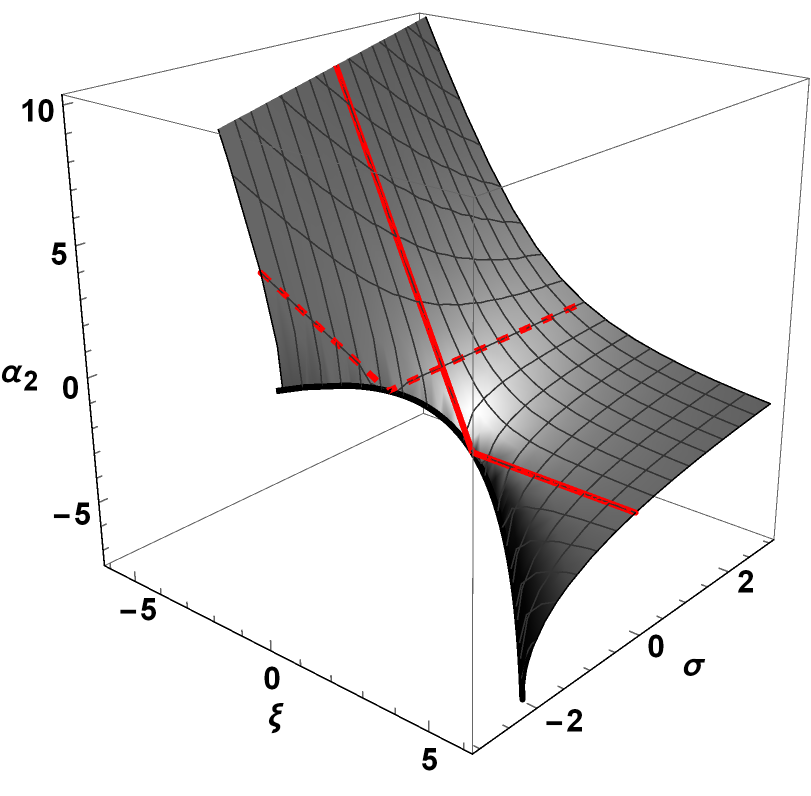}
\caption{$\alpha_1$ (top) and $\alpha_2$ (bottom) in a neighbourhood of $\xi=0$ (dashed red curve) and $\sigma=0$ (solid red curve) for $w=0$. The solid black curve refers to the case $\Delta=0$.
}
\label{fig1}
\end{figure}

For completeness we remark that in the case $\xi=0$ we obtain either $\alpha_1=-3(1+\sigma)$ and $\alpha_2=0$, or $\alpha_1=0$ and $\alpha_2=-3(1+\sigma)$.
In Figure \ref{fig1} we display $\alpha_1$ (top) and $\alpha_2$ (bottom) in a neighbourhood of $\xi=\sigma=0$.
Here we see that the noninteracting configuration pinch into the point $\xi=\sigma=0$
(the intersections of dashed and solid red curves). 

It is easy to see
that in the domain $\alpha_1<0$ and $\alpha_2<0$, 
\begin{eqnarray}
\label{eqa111}
\lim_{a\rightarrow \infty}\rho =0,~~\lim_{a\rightarrow \infty}V =V_\Lambda.
\end{eqnarray}
From the Friedmann equation (\ref{eqm9}) it is then easy to see that when $a\rightarrow \infty$ we obtain one attractor (stable) de Sitter configuration
and one repeller (unstable) de Sitter configuration.

Let us now consider the first integral given by the Friedmann equation (\ref{eqm9}).
Substituting (\ref{eqa81n}) and (\ref{eqa82n}) in (\ref{eqm9}) we obtain
\begin{eqnarray}
\label{eqa12q}
H^2+U(a)=\kappa^2 \frac{V_\Lambda}{3},
\end{eqnarray}
where the potential $U(a)$ is given by
\begin{eqnarray}
\label{eqa13}
U(a)=\frac{k}{a^2}-\frac{\kappa^2}{3}[(E_1+\lambda_1) a^{\alpha_1}+(E_2+\lambda_2) a^{\alpha_2}].
\end{eqnarray}
Taking into account that
\begin{eqnarray}
\label{eqads}
\dot{a}=Ha
\end{eqnarray}
together with the time derivative of (\ref{eqa12q})
\begin{eqnarray}
\label{eqa14}
\dot{H}&=&-\frac{a}{2}\frac{dU}{da},
\end{eqnarray}
we now have a two-dimensional dynamical system for $a$ and $H$.
%
We define $P_c=(a_c, H_c)$ to be stationary solutions -- fixed points --
of equations (\ref{eqads}) and (\ref{eqa14}). 
From a direct inspection of (\ref{eqads}) we see that our dynamical system might support fixed points
with $a_c=0$. However, equations (\ref{eqads}) and (\ref{eqa14}) must be subject to the Friedmann
constraint (\ref{eqa12q}) which may be singular for $a_c=0$. 
As we are interested in nonsingular
configurations
we will not take into account such fixed points. On the other hand, if $a_c$ is defined as solutions of ${dU}/{da}|_{a_c}\equiv 0$ with $H_c\equiv H|_{a_c}=0$, we see that such fixed points $P_c=(a_c, 0)$ are related 
to the extrema of the potential $U(a)$. 
These points, if they exist, represent Einstein static models.

Expanding (\ref{eqads}) and (\ref{eqa14}) in a neighbourhood
of $P_c$, we obtain
\begin{eqnarray}
\label{eqa2}
\dot{\Phi}_i=L_{ij}|_{P_c}\Phi_j
\end{eqnarray}
where $\Phi_i$ is the $2$-vector 
\begin{eqnarray}
\label{eqa3}
 \Phi_{i}
\rightarrow
   \begin{bmatrix}
   a-a_c  \\
   H-H_c       
   \end{bmatrix}.
\end{eqnarray}
and $L_{ij}\equiv \partial \dot{\Phi}_i/\partial \Phi_j$ is the Jacobian of the dynamical system.
That is,
\begin{eqnarray}
\label{eqa3ai}
 L_{ij} (a, H)
\rightarrow
   \begin{bmatrix}
   H             & a                    \\
-\frac{1}{2}\Big(\frac{dU}{da}+a\frac{d^2U}{da^2}\Big)  &  0      
   \end{bmatrix}.
\end{eqnarray}
It is then easy to see that the eigenvalues of $L_{ij}$ evaluated at $P_c$ are given by
\begin{eqnarray}
\label{eigen1}
\gamma_{\pm}=\pm a_c\sqrt{-\frac{1}{2}\frac{d^2U}{da^2}\Big|_{a_c}}.
\end{eqnarray}
This is a crucial result which dictates the stability of the dynamics in a neighbourhood of fixed points
in the phase space. In fact, if ${d^2U}/{da^2}|_{a_c}>0$ we obtain that $\gamma_\pm$ are pure imaginary so that the corresponding fixed point is a center. In this case the linearization theorem \cite{AP} fails to establish the stability (or otherwise) of the fixed point, but numerical integration of the equations confirms that this fixed point is a centre, surrounded by cyclic trajectories. 
For ${d^2U}/{da^2}|_{a_c}<0$ on the other hand, $\gamma_\pm$ are real and the corresponding fixed point is a saddle.
We remark again that both fixed points represent Einstein static models; however the saddle one is similar to the original Einstein model due to a cosmological constant term, the first is a centre like the one appearing in loop-quantum cosmology \cite{Parisi:2007kv} or in some models with a quadratic equation of state \cite{Ananda:2005xp,Ananda:2006gf,Ganguly:2019llh}.
Considering our solution (\ref{eqa81n}) and (\ref{eqa82n}),
we are interested in the case where the evolution of the energy density can mimic a non-interacting cosmology, but includes a correction term which might also lead to a bounce in the very early universe while preserving the weak energy condition $\rho+V>0$.
It can be easily shown that $\sigma$ vanishes if one fixes $\alpha_1=-3(1+w)$ or $\alpha_2=-3(1+w)$.
Therefore, to simplify our analysis, in the next section we will focus on the case $\xi\neq 0$ and $\sigma=0$. In the following we will also show that analogous models can be built for the case
$\xi= 0$ and $\sigma\neq 0$ as long as an additional  perfect fluid is included in order to construct a nonsingular model with a noninteracting matter-dominated era.

\subsection{\boldmath$\xi\neq 0$ and \boldmath$\sigma=0$}

In this case {from \eqref{delta} we choose $\Delta=1+w -\xi$, then from \eqref{eqa91}-\eqref{eqa92} we have $\alpha_1=-3\xi$, $\alpha_2=-3(1+w)$ and}  the solutions  (\ref{eqa81n}) and (\ref{eqa82n}) reduce to
\begin{eqnarray}
\label{eqa01}
\rho&=&\frac{E_1}{a^{3(1+w)}}+\frac{E_2}{a^{3\xi}},\\
\label{eqa02}
V&=&V_\Lambda+\frac{\lambda_2}{a^{3\xi}}.
\end{eqnarray}
$\lambda_1=0$ by the virtue of (\ref{eqa101}) and from (\ref{eqa102}) 
\begin{eqnarray}
\label{el1}
\lambda_2=\frac{E_2}{\xi}(1+w-\xi).
\end{eqnarray}
In (\ref{eqa01}) we identify the first term as the conventional (non-interacting) component of the fluid density for $E_1>0$.
The second terms of (\ref{eqa01}) and (\ref{eqa02}) on the other hand, are exotic terms due to vacuum interaction which combined in the Friedmann
equation may give a bounce for $E_2<0$, as discussed below.

We now examine the Friedmann equation (\ref{eqa12q}) in order to determine an appropriate domain for the parameters $w$ and $\xi$.
The potential (\ref{eqa13}) can be written as 
\begin{eqnarray}
\label{eqa05}
U(a)=\frac{k}{a^2}-\frac{\kappa^2}{3}\Big[\frac{E_1}{a^{3(1+w)}}+\frac{E_2(1+w)}{\xi a^{3\xi}}   \Big].
\end{eqnarray}
For $0\leq w \leq 1$, from the above we see that a sufficient 
condition to obtain a nonsingular bounce for $E_1>0$ is given by
\begin{eqnarray}
\label{eqa06}
\frac{E_2(1+w)}{\xi} <0~~{\rm and}~~ \xi >(1+w).
\end{eqnarray}
In fact, in this case an infinite potential barrier avoids the classical singularity
found in the noninteracting case. In the following we shall restrict ourselves to such configurations.

To give a numerical illustration, from now on we will consider the case of a nonrelativistic 
perfect fluid so that $w=0$, and we also fix $\xi=4/3$. In this case, the potential (\ref{eqa05}) of Friedmann equation turns into
\begin{eqnarray}
\label{eqa07}
U(a)=-H_0^2\Big(\frac{\Omega_{k0}}{a^2}+\frac{\Omega_{m0}}{a^3}+\frac{\Omega_{I0}}{a^4}\Big),
\end{eqnarray}
where 
\begin{eqnarray}
\label{eqa08}
\Omega_{k0}=-\frac{k}{H^2_0},~~\Omega_{m0}=\frac{\kappa^2 E_1}{3H^2_0},
~~\Omega_{I0}=\frac{\kappa^2 E_2}{4H^2_0}.
\end{eqnarray}
Here and in the remainder of this section we use the  normalization for the scale factor such that $a_0=1$ at present.

Restricting ourselves to the case $\Omega_{I0}<0$ (or $E_2<0$), it can be easily seen that 
\begin{eqnarray}
\label{eqa09}
\lim_{a\rightarrow 0^+} U(a)=+\infty,
\end{eqnarray}
so that an infinite potential barrier avoids the classical singularity.
For $\Omega_{k0}\neq 0$ it can be shown that the potential (\ref{eqa07}) has at most two extrema, $a_{c\pm}>0$, connected to fixed points of Eqs.~(\ref{eqads}) and~(\ref{eqa14}). As mentioned above, such fixed points are given by $P_c=(a_{c\pm},0)$
where in this case
\begin{eqnarray}
\label{eqa010}
a_{c\pm} = \frac{-3\Omega_{m0}\pm \sqrt{-32\Omega_{I0}\Omega_{k0}+9\Omega^2_{m0}}}{4\Omega_{k0}}.
\end{eqnarray}
The condition for the two extrema to be real and positive is
\begin{eqnarray}
\label{eqa011}
\bar{\Omega}_{k0}\equiv\frac{9\Omega_{m0}^2}{32\Omega_{I0}} < \Omega_{k0} <0
\,.
\end{eqnarray}
%
No extrema exist for $\Omega_{k0}<\bar{\Omega}_{k0}$, while for $\Omega_{k0}\geq 0$ (spatially flat or open models) there is only one positive extremum, $a_{c+}>0$, and the potential (\ref{eqa07}) has one global minimum in the domain $a>0$.
%
For all these models we see that bouncing models exist due to the exotic interaction term, $\Omega_{I0}<0$, which provides a potential barrier in $U(a)$. 

%
%
%

To examine the structure of the phase-space for nonsingular models 
let us consider the equations (\ref{eqads}) and~(\ref{eqa14})
together with the potential (\ref{eqa07}). For this case the linearization matrix (\ref{eqa3}) reads
\begin{eqnarray}
\label{eqa3b}
 L_{ij} (a, H)
 \rightarrow
  \begin{bmatrix}
        H           & a                   \\
        \frac{H_0^2}{a^3}\Big(2\Omega_{k0}+\frac{9\Omega_{m0}}{2a}+\frac{8\Omega_{I0}}{a^2}       \Big)          &  0       \\
   \end{bmatrix},
\end{eqnarray}
%
and its eigenvalues 
evaluated at $P_c=(a_{c}, 0)$ 
are
\begin{eqnarray}
\label{evnew}
\gamma_{\pm}=\pm\frac{H_0}{a_{c}^2}\sqrt{2(4\Omega_{I0}+a^2_c\Omega_{k0})+9a_c\Omega_{m0}/2}.
\end{eqnarray}

\begin{figure}
\includegraphics[width=8cm,height=5cm]{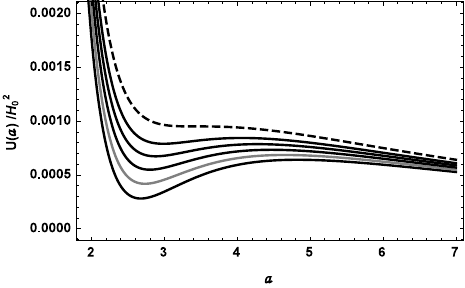}
\includegraphics[width=8cm,height=5cm]{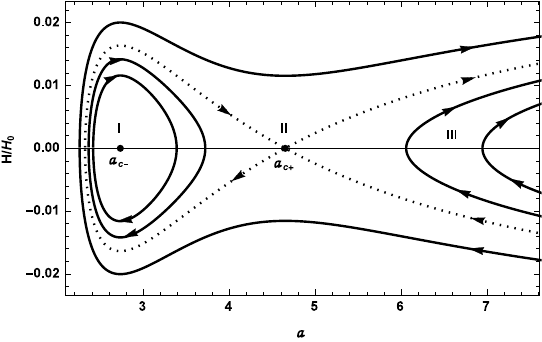}
\caption{
The potential $U(a)$ (top panel) for $\Omega_{m0}=0.31$ and $\Omega_{I0}=-0.40$ for different values of $\Omega_{k0}<0$. The top (dashed) curve corresponds to the upper
limit $\Omega_{k0}=\bar{\Omega}_{k0}\simeq -0.067$. The phase portrait is displayed on the bottom panel where for the purposes of illustration we fix $\Omega_{k0}=-0.063$ (grey curve in top panel). The fixed points, $a_{c-}\simeq 2.730$ (center) and $a_{c+}\simeq 4.650$ (saddle),  correspond to the extrema of the potential (\ref{eqa010}).
For $V_\Lambda< 3U(a_{c+})/\kappa^2$ we obtain
cylic universes in a finite neighbourhood (region I) of $a_{c-}$. One-bounce orbits (in region III) can also be obtained for this domain of $V_{\Lambda}$.
For $V_{\Lambda}=3U(a_{c+})/\kappa^2$, a separatrix emerges from the saddle fixed point $a_{c+}$ which defines an escape to the de Sitter attractor at infinity.
Finally, for $V_{\Lambda}> 3U(a_{c+})/\kappa^2$ we obtain one-bounce orbits in regions II.} %
%
\label{fig2}
\end{figure}

\begin{figure}
\includegraphics[width=7.3cm,height=7cm]{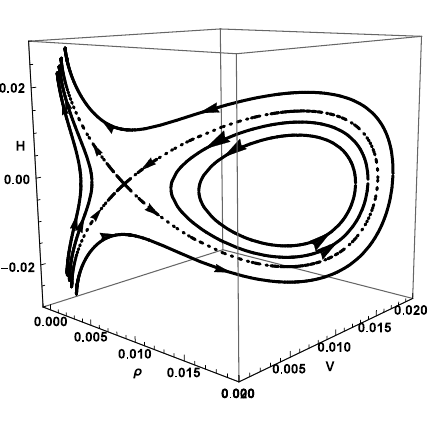}
\caption{The phase portrait of the full phase space $(\rho, V, H)$
in units $\kappa^2=H_0=1$.
The corresponding fixed points with $H_c=0$ are given by
$(\rho_{c-} \simeq  0.0168,~~V_{c-}\simeq 0.0092)$ and
$(\rho_{c+}\simeq  0.0058,~~V_{c+}\simeq 0.0029)$, 
in units of $\kappa^2 =H_0=1$.
}
\label{fig3}
\end{figure}

We illustrate the behaviour of the potential (\ref{eqa07}) in the top panel of Fig. \ref{fig2}, focusing on the case of closed models ($\Omega_{k0}<0$), fixing the parameters
$\Omega_{m0}=0.31$ and $\Omega_{I0}=-0.40$. In the bottom panel we illustrate several orbits in the plane $(a, H/H_0)$
for $\Omega_{k0}=-0.063$ \footnote{We note that current Planck data allow positive spatial curvature, $\Omega_{0k}=-0.001\pm0.002$~\cite{Planck:2018vyg}, and there are also some arguments favouring a nonvanishing $3$-curvature~\cite{DiValentino:2019qzk,Handley:2019tkm}.}.
Orbits in region~II of the bottom panel of Fig. \ref{fig2} are of physical interest in the sense that they show a transition from an early 
phase at high energy
to a decelerated matter era together with a graceful exit to a late-time accelerated regime.
We see that the eigenvalues of $L_{ij}(a_{c-}, 0)$ are pure imaginary: then the numerical integration of the equation confirms that 
$a_{c-}$ is a center fixed point representing a stable Einstein static universe.
On the other hand, the eigenvalues of $L_{ij}(a_{c+}, 0)$ are real so that $a_{c+}$ is a saddle (an unstable Einstein static universe).
In Fig. \ref{fig3} we show the phase portrait of the full phase space $(\rho, V, H)$. 
\begin{figure}
\includegraphics[width=8cm,height=5cm]{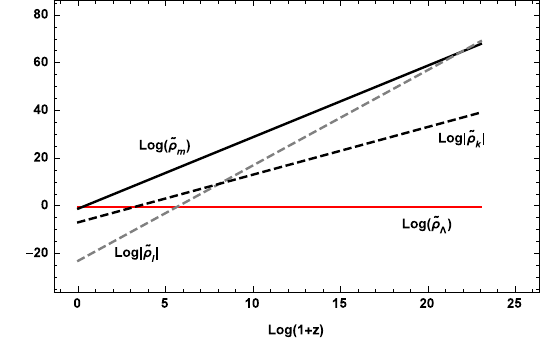}
\caption{The logarithm of the dimensionless densities $\tilde{\rho}_m\equiv \Omega_{m0}/a^3$, $\tilde{\rho}_k\equiv \Omega_{k0}/a^2$ and $\tilde{\rho}_I\equiv \Omega_{I0}/a^4$ as a function of the redshift $1+z\equiv a^{-1}$.
The curvature density on the other hand is pushed towards $10^{17}$ at the bounce. For comparison we also show (in red) the logarithm of $\tilde{\rho}_\Lambda\equiv \Omega_{\Lambda 0}$.
}
\label{fig4}
\end{figure}
%

In order to illustrate a more realistic scenario, we next set 
\begin{eqnarray}
\label{vac}
\Omega_{\Lambda 0} \equiv \frac{\kappa^2 V_\Lambda}{3H_0^2}\simeq 0.68. 
\end{eqnarray}
As Planck data~\cite{Planck:2018vyg} do leave some
room for curvature we fix $\Omega_{k0}=-0.001$ for illustration. 
Assuming again $\Omega_{m0}=0.31$,
the remaining task is to determine suitable values for $\Omega_{I0}$.

Considering the evolution
of quantum cosmological perturbations, it has been shown~\cite{Peter:2006hx, Pinto-Neto:2009kyd} that in order to obtain primordial perturbations from a bounce compatible with CMB data, one must satisfy the condition $R^{-1/2} \gtrsim 10^3 \times l_p$, where 
$R$ is the Ricci scalar and $l_p$ the Planck length. On the other
hand, in order to reproduce features of a conventional hot big bang cosmology at high redshift such as the cosmic neutrino background \cite{Lesgourgues:2006nd}, the bounce must occur at a redshift greater than $z\simeq 10^{10}$.
Bearing such considerations in mind, we will assume that the physical domain of the bounce parameter is bounded by
\begin{eqnarray}
\label{eqa013}
-10^{-38}\lesssim \Omega_{I0} \lesssim -10^{-10}.
\end{eqnarray}
For every value of $\Omega_{I0}$ in this range, over $28$ orders of magnitude,
we obtain a nonsingular model corresponding to orbits in the phase space analogous to those in
region II of the bottom panel of Figure~2. 

To illustrate a particular case let us fix $\Omega_{I0}= - 10^{-10}$.
For this simple configuration we obtain
$a_{c-}\simeq 1.07\times 10^{-10},~~a_{c+}\simeq 465$.
However, contrary to what one might expect, the transition from an accelerated early phase to a decelerated era in which
matter dominates does not take place at $a_{c-}$. Equivalently, the transition from such a decelerated era 
to a de Sitter regime does not take place at $a_{c+}$. In fact, the extrema of the potential $U(a)$ correspond to stationary solutions of the equation $\dot{H}=0$. To get a proper evaluation of transitions phases we actually need to solve the equation
for $\ddot{a}$. To this end, let us rewrite the first Friedmann equation as
\begin{eqnarray}
\label{eqfn27}
\dot{a}^2+W(a)=0,
\end{eqnarray} 
where $W(a) \equiv a^2 [U(a)-\kappa^2 V_\Lambda/3] $. Therefore
\begin{eqnarray}
\label{eqpw}
\ddot{a}=-\frac{1}{2}\frac{dW}{da}.
\end{eqnarray}
Defining $a_I$ as the transition from an accelerated early universe to a decelerated matter era and 
$a_{II}$ as the transition from such a decelerated era 
to a de Sitter regime, it can be easily shown that
\begin{eqnarray}
\label{trs}
a_I\simeq 1.61 \times 10^{-10},~~a_{II}\simeq 0.61.
\end{eqnarray}
For example, according to observations~\cite{SupernovaSearchTeam:1998fmf}, a transition from a matter dominated era to a de Sitter regime 
should take place at a redshift $z_{II} \simeq 0.426^{+0.27}_{-0.089}$ (or, $a_{II} \simeq 0.7^{+0.04}_{-0.11}$). 
From (\ref{trs}) we see that the $a_{II}$ given in Eq.~(\ref{trs}) lies within this domain.  
%

In Fig. \ref{fig4} we show the behaviour of $\Omega_{m0}/a^3$, $\Omega_{k0}/a^2$ and $\Omega_{I0}/a^4$ as a function of the redshift $z\equiv a^{-1}-1$.
At the bounce ($z_b\simeq 10^{10}$), the matter and interaction densities are of the order of $10^{29}$.
The curvature density on the other hand is pushed towards $10^{17}$ at the bounce. In this figure it is also shown that the vacuum parameter $V_\Lambda$ plays a significant role only at late times near $z\simeq 0$.

Last but not least, it can be shown that for $w=0$ and $\xi=4/3$, the bounce scale $a_b$ 
is typically of the order of $|\Omega_I|$. Therefore, for the domain (\ref{eqa013}) -- together with the chosen parameters 
$\Omega_{m0}=0.31$ and $\Omega_{\Lambda 0}=0.68$ -- it is 
easy to show that the weak energy condition $\rho+V>0$ is automatically satisfied.

\subsection{\boldmath$\xi= 0$ and \boldmath$\sigma\neq 0$}

In this case, the above solutions (\ref{eqa81n}) and (\ref{eqa82n}) reduce to
\begin{eqnarray}
\label{eqb01}
\rho &=&\frac{E_1}{a^{3(1+w +\sigma)}},\\
\label{eqb02}
V &=&V_\Lambda- \frac{\sigma E_1}{(1+w+\sigma)a^{3(1+w +\sigma)}},
\end{eqnarray}
where we have used (\ref{vinc1})-(\ref{vinc2}).
In order to seek nonsingular configurations we note that
\begin{eqnarray}
\label{eqb03}
\rho+ V\equiv V_\Lambda+\frac{ E_1(1+w)}{(1+w+\sigma)
a^{3(1+w+\sigma)}}.
\end{eqnarray}
Taking a glance at the first Friedmann equation (\ref{eqm9}), 
we see that nonsingular models might be obtained for $1+w+\sigma>0$ as long as 
$(1+w)E_1<0$. 
In this sense, nonsingular models analogous to that of the preceding subsection can be built.
However, if we set $w=0$, as we considered in the previous subsection, then a conventional matter-dominated era, with $\rho\propto a^{-3}$ cannot be achieved for $\sigma\neq0$, unless an additional noninteracting pressureless fluid is included.


\section{The Nonlinear case \boldmath$Q=Q_2$}

Substituting the nonlinear interaction (\ref{eqm112}) into the continuity equations~(\ref{eqm71}) and (\ref{eqm72}) gives
\begin{eqnarray}
\label{nc1}
\dot{\rho}&=&-3H\rho\Big[1+w+\chi \Big(1-\frac{V}{V_\Lambda}\Big)\Big],\\
\label{nc2}
\dot{V}&=&3\chi H\rho \Big(1-\frac{V}{V_\Lambda}\Big).
\end{eqnarray}
In addition, these equations are coupled with the Raychaudhuri equation (\ref{eqm10})
\begin{eqnarray}
\label{Ray}
\dot{H} =-H^2 -\frac{\kappa^2}{6}[\rho(1+3w)-2V].
\end{eqnarray}
We thus have three coupled equations describing the dynamics of the three-dimensional system $(\rho,V,H)$. 

\subsection{Vacuum sub-manifold}

First we note that there is an invariant sub-manifold corresponding to vacuum cosmologies where $\rho=0$. In this case the interaction $Q_2$ in (\ref{eqm112}) vanishes and we have from (\ref{nc2}) that $\dot{V}=0$, hence the vacuum energy is given by an integration constant, $V=V_{dS}=$constant. The only remaining dynamical equation is the Raychaudhuri equation (\ref{Ray}) which reduces to 
\begin{eqnarray}
\label{Ray-vacuum}
\dot{H} =-H^2 +\frac{\kappa^2}{3}V_{dS} \,.
\end{eqnarray}
This is just the evolution equation of the de Sitter spacetime in its  FLRW representation, which in general includes curvature. 
The fixed points on this sub-manifold correspond to the spatially flat de Sitter model $(\rho,V,H)=(0,V_{dS},H_{dS})$, where
\begin{eqnarray}
\label{dSHc}
H_{dS}=\pm\kappa\sqrt{\frac{V_{dS}}{3}}
\end{eqnarray}
are the contracting and expanding versions of the model, and $\kappa^2 V_{dS}$ is a cosmological constant. We remark that the closed model evolves between the two with a bounce, according to \eqref{Ray-vacuum}.

\subsection{Non-interacting sub-manifold}
\label{sec:NIsub}

\begin{figure}
\includegraphics[width=8cm
]{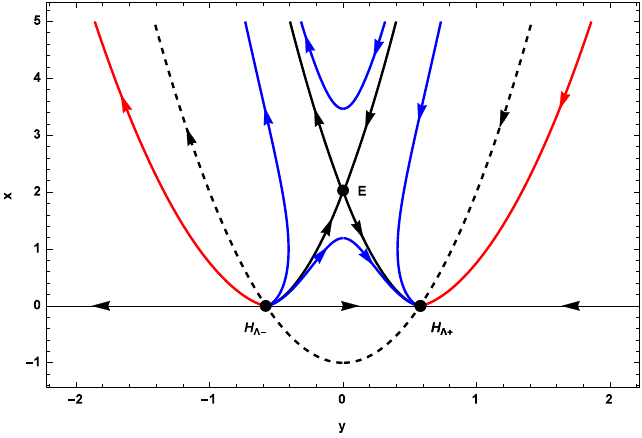}
\caption{The phase plane for the non-interacting sub-manifold $V=V_\Lambda$ discussed in subsection \ref{sec:NIsub}. We define $x=\rho/V_\Lambda$ and $y=H/\sqrt{V_\Lambda}$.
The black dots on the $x=0$ line denote the de Sitter fixed points. The black dot at $y=0$ denotes the Einstein fixed point. 
Separatrices are shown in black.
The black dashed parabola corresponds to flat models ($k=0$). 
All trajectories inside the parabola (blue orbits) are positively curved ($k>0$). All trajectories outside the parabola (red orbits) are related to models of negative curvature ($k<0$). The lower blue trajectory  connecting the two de Sitter fixed points is representative of models that collapse to a bounce at $y=0$ then re-expand.}
\label{figps}
\end{figure}

Second, we note that if 
$V=V_\Lambda$, then \eqref{nc2} tells us  that $V=V_\Lambda$ at all times, i.e., $V_\Lambda$ is a cosmological constant, and $V=V_\Lambda$ defines another invariant submanifold in phase space. In this case there is no interaction and we have a conventional non-interacting cosmology ($\Lambda$CDM when $w=0$) described by the two-dimensional dynamical system 
\begin{eqnarray}
\label{nc1-LCDM}
\dot{\rho}&=&-3(1+w)H\rho ,\\
\label{Ray-LCDM}
\dot{H} &=& -H^2 -\frac{\kappa^2}{6}[\rho(1+3w)-2V_\Lambda].
\end{eqnarray}
The $(\rho,H)$ phase plane for the non-interacting dynamics with $V=V_\Lambda$ is plotted in Fig. \ref{figps}. 
From \eqref{nc1-LCDM}-\eqref{Ray-LCDM} we see that this dynamical system has three fixed points, assuming $w>-1/3$. The first appears at $H=0$ and we may call it an Einstein saddle, as it represents an Einstein static model~\cite{Ananda:2005xp,Ananda:2006gf,Parisi:2007kv,Maier:2013hr} with 
\begin{eqnarray}\label{EP}
\rho_E = \frac{2V_E}{1+3w}\,.
\end{eqnarray}
The other two fixed points on this non-interacting sub-manifold correspond to $\rho=0$, i.e.\ where this  sub-manifold intersects the vacuum sub-manifold. These two fixed points 
are therefore given by $(\rho,V,H)=(0,V_\Lambda,H_{\Lambda\pm})$, where from Eq.~(\ref{Ray-LCDM})
\begin{eqnarray}
\label{dsn}
H_{dS}=H_{\Lambda\pm}=\pm\kappa\sqrt{\frac{V_\Lambda}{3}}\,.
\end{eqnarray} 
Thus they represent de Sitter models, one contracting and one expanding, with zero spatially curvature. The line $\rho=0$ between the two de Sitter fixed points in Fig.~\ref{figps} represents a de Sitter spacetime with positively curved space, and outside the points it is de Sitter spacetime with negatively curved space.

\subsection{$\Sigma$ sub-manifolds}

More generally, for $\rho\neq0$ and $V\neq V_\Lambda$,
the integration of (\ref{nc1}) and (\ref{nc2}) is a rather more involved task than for the linear case
due to its intrinsic nonlinearity. However, an alternative method may be employed to reduce the number 
of degrees of freedom of the dynamics and identify other invariant sub-manifolds. In fact,
we can eliminate $H$ from equations (\ref{nc1}) and (\ref{nc2}) to obtain,
\begin{eqnarray}
\label{nca3}
\dot{\rho}+\Big[1+\frac{1+w}{\chi (1-V/V_\Lambda)}\Big] \dot{V}=0 \,.
\end{eqnarray}
A direct integration of (\ref{nca3}) gives
\begin{eqnarray}
\label{nca4}
\rho =\rho_\ast-(V-V_\ast)\!
+\frac{V_\Lambda(1+w)}{2\chi}\ln\Big[{\Big(\frac{1-V/V_\Lambda}{1-V_\ast/V_\Lambda}\Big)^2}\Big],
\end{eqnarray}
where $\rho_\ast$ and $V_\ast$ correspond to the initial values at $t=t_\ast$.
In general,  every pair of initial conditions $\rho_\ast$ and $V_\ast$ (and set of parameters values) defines a 2-dimensional surface 
$\Sigma$
in the 3-dimensional phase space, $(\rho, V, H)$, 
%
characterised by the first integral
\begin{eqnarray}\label{FI}
{\mathcal K}_\Sigma=\rho +V - \frac{V_\Lambda(1+w)}{2\chi} \ln\Big[\Big(1-\frac{V}{V_\Lambda}\Big)^2\Big]\; .
\end{eqnarray}
%

\begin{figure}
\includegraphics[width=8cm
]{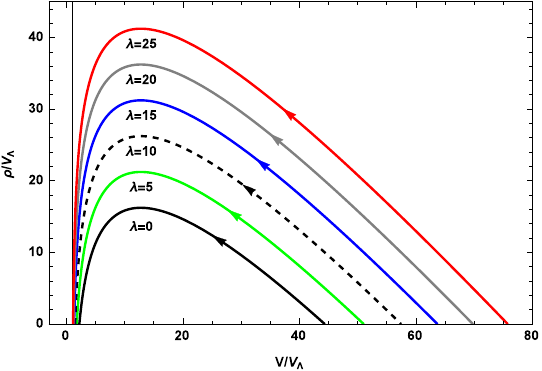}
\includegraphics[width=8cm
]{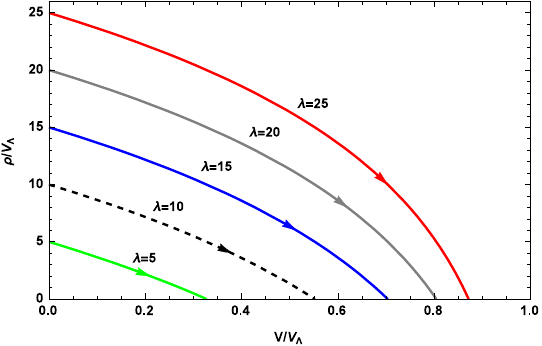}
\caption{
The ($\rho,V$) phase plane for $\chi=0.085$, where each curve corresponds to a different value of the first integral (\ref{FI}) 
$\lambda \equiv {\mathcal K}_\Sigma /V_\Lambda$.  Arrows indicate the evolution with time during an expansion phase for $V_\Lambda>0$. {\it Top panel:} only trajectories for $V/V_\Lambda > 1$ are shown for clarity. 
The vertical line at $V/V_\Lambda = 1$ corresponds to the non-interacting sub-manifold. {\it Bottom panel:}  zoom-in of the phase plane for $0< V/V_\Lambda < 1$. }
\label{fig7}
\end{figure}

This allows us to depict the phase plane ($\rho,V$) directly, as shown in  Figure~\ref{fig7}, where each line corresponds to a trajectory, or a union of trajectories.
We study the motion on the surface $\Sigma$ by
substituting (\ref{nca4}) into (\ref{nc2}) and (\ref{Ray}) to eliminate $\rho$, obtaining
\begin{eqnarray}
\label{nca5}
\nonumber
\dot{V}&=&{3 H}\Big(1-\frac{V}{V_\Lambda}\Big)\Big\{\chi (\rho_\ast+V_\ast-V)
\\
&&
+\frac{V_\Lambda(1+w)}{2} \ln{\Big[\Big(\frac{1-V/V_\Lambda}{1-V_\ast/V_\Lambda}\Big)^2\Big]}\Big\},
\\
%
%
\label{nca6}
\nonumber
\dot{H}&=&-H^2+\frac{\kappa^2}{6}\Big\{3V(1+w)-(\rho_\ast+V_\ast)(1+3w)\\
&&
-\frac{V_\Lambda(1+w)(1+3w)}{2\chi}\ln{\Big[\Big(\frac{1-V/V_\Lambda}{1-V_\ast/V_\Lambda}\Big)^2\Big]}\Big\},
\end{eqnarray}
%
after integrating the differential equations (\ref{nc1})--(\ref{Ray}).
In the following we shall examine the fixed points of (\ref{nca5})--(\ref{nca6}).


In addition to the previously identified de Sitter fixed points where $\rho=0$, we see that the dynamical system (\ref{nc1})--(\ref{Ray}) also admits fixed points where $H=0$. Again, we identify these as Einstein static universes~\cite{Ananda:2005xp,Ananda:2006gf,Parisi:2007kv,Maier:2013hr}. Indeed, the condition $H=0$ is enough to ensure both $\rho$ and $V$ are constants; thus we denote this fixed point via $(\rho,V,H)=(\rho_E,V_E,0)$, where, again assuming $w>-1/3$, Eq.~\eqref{Ray} gives the  relation \eqref{EP} between the matter and vacuum energy densities.

Substituting the constraint (\ref{EP}) into the first integral (\ref{FI}) we have
\begin{eqnarray}
\label{FIE}
{\mathcal K}_{\Sigma}=(1+w) \left\{ \frac{3V_E}{1+3w} - \frac{V_\Lambda}{2\chi} \ln\Big[\Big(1-\frac{V_E}{V_\Lambda}\Big)^2\Big] \right\} \,.
\end{eqnarray}
Equation (\ref{FIE}) is a transcendental equation which cannot in general be analytically solved for $V_E$.

Expanding (\ref{nca5})--(\ref{nca6}) in a neighbourhood 
of general fixed points we obtain
\begin{eqnarray}
\label{nc3}
\dot{\Psi}_i=L_{ij}|_{{P}_{c}}\Psi_j,
\end{eqnarray}
where $\Psi_i$ is the $2$-vector
\begin{eqnarray}
\label{psivec}
\Psi_{i}
\rightarrow
\begin{pmatrix}
V-V_{c}\\
H-H_{c}   
\end{pmatrix}   
\end{eqnarray}
and  $L_{ij}\equiv \partial \dot{\Psi}_i/\partial \Psi_j$.
It is then easy to verify that the eigenvalues of $L_{ij}$ evaluated at the Einstein static fixed points
are given by
\begin{eqnarray}
\label{ev}
\tilde {\gamma}_{\pm}=\pm \kappa 
\sqrt{\frac{(1+w)V_E\left[(1+3w+3\chi)V_\Lambda-3\chi V_E\right]}{(1+3w)V_\Lambda}}
\,.
\end{eqnarray}
Thus we see that the Einstein static fixed points are either saddle points (real eigenvalues) or centers (imaginary eigenvalues).

\subsection{Non-singular solutions}

The main configurations of interest which will guide 
our analysis from now on are those connected to nonsingular models in the sense that
after a bounce the universe is driven towards a decelerated phase together with a grateful exit to 
a de Sitter attractor, similar to the behaviour seen in Figure~\ref{fig2}.
Such configurations may be obtained as long as two different fixed points, a saddle point and a center, are present,
given that the presence of a de Sitter attractor is already guaranteed 
by (\ref{dsn}).
In order to show specific examples we will fix $w=0$ in this sub-section.

\begin{figure}
\includegraphics[width=8.5cm
]{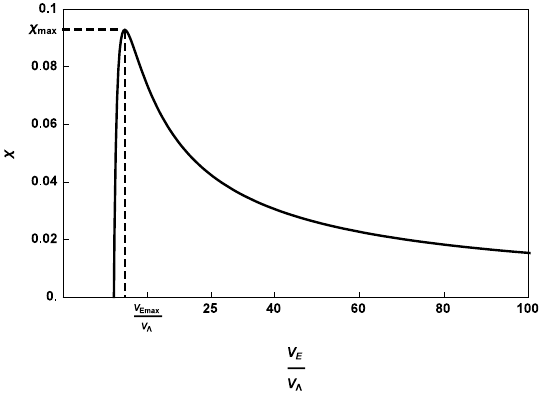}
\caption{$\chi$ as a function of $V_E/V_\Lambda$ according to (\ref{chi0}) in the domain $V_E>2V_\Lambda$.
For the purpose of illustration we fixed ${\mathcal K}_\Sigma=0$.
The global extrema in the above domain is located at
$(V_{Emax}/V_\Lambda, \chi_{max})$ 
and satisfies equation (\ref{ext}). Here we see that for a fixed value of $\chi$ in the domain $0<\chi<\chi_{max}$ there are two
fixed points (connected to two distinct values of $V_E/V_\Lambda$).
}
\label{fig8}
\end{figure}

We shall proceed by searching for a proper
domain of $\chi$ in which such non-singular models may be obtained.  
From equation (\ref{FIE}) 
one may write $\chi$ as a function of 
$V_E$ as
\begin{eqnarray}
\label{chi0}
\chi=\frac{V_\Lambda}{2(3 V_E- {\mathcal K}_\Sigma)}\ln\Big[\left({1-\frac{V_E}{V_\Lambda}}\right)^2\Big].
\end{eqnarray}
From (\ref{chi0}) it is easy to see that $\chi\rightarrow 0$ as $V_E/V_\Lambda\rightarrow +\infty$.
On the other hand, $\chi$ has a root at $V_E=2V_\Lambda$ and diverges as $V_E\rightarrow V_\Lambda$.
Therefore, in order to simplify our analysis we are going to restrict ourselves to the 
case $V_E>V_\Lambda$.
%
%
In this domain it is then easy to show that $\chi$ as a function of $V_E$ has a global extrema -- located at $V_{Emax}$ --
which satisfies the relation
\begin{eqnarray}
\label{ext}
\frac{V_{Emax}}{V_\Lambda} = \frac{1+3\chi_{max}}{3\chi_{max}}.
\end{eqnarray}
In Fig.~\ref{fig8} we show the behaviour of $\chi$ as a function of $V_E$ -- for ${\mathcal K}_\Sigma=0$ -- in the domain $V_E>V_\Lambda$.
For the purpose of illustration, in the following we shall restrict to this case. In Fig. \ref{fig7} (bottom panel) we display the initial conditions $(\rho_\ast, V_\ast)$ connected to the first integral (\ref{FI}) with ${\mathcal K}_\Sigma=0$.

%
%
\begin{figure}
\includegraphics[width=8cm,height=5cm
]{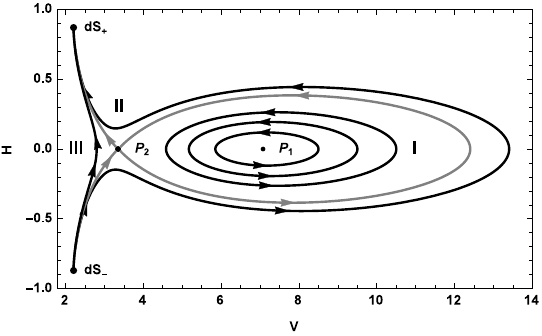}  \includegraphics[width=8cm,height=5cm]{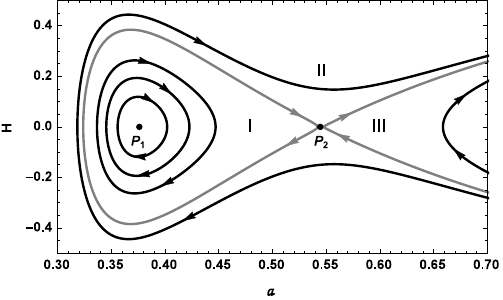}
\caption{{\it Top panel:} The phase portrait for $(V,H)$ with non-linear coupling parameter $\chi=0.085$, shown for several different initial conditions with first integral ${\mathcal K}_\Sigma=0$.
Here we have also fixed $V_\Lambda=\kappa^2=1$. A separatrix (grey curve) emerges from the saddle fixed point, $P_2$, dividing the phase space in two distinct regions; perpetually-cycling orbits  (region I in the neighbourhood of the center fixed point $P_1$) and one-bounce orbits (in regions II and III).
Nonsingular orbits in region II are of physical interest in the sense that they show a transition from a decelerated phase to late-time acceleration. 
{\it Bottom panel:} The corresponding orbits for $(a, H)$ with the same initial conditions for the case $k>0$. }
\label{fig10}
\end{figure}
\begin{figure}
\includegraphics[width=10cm
]{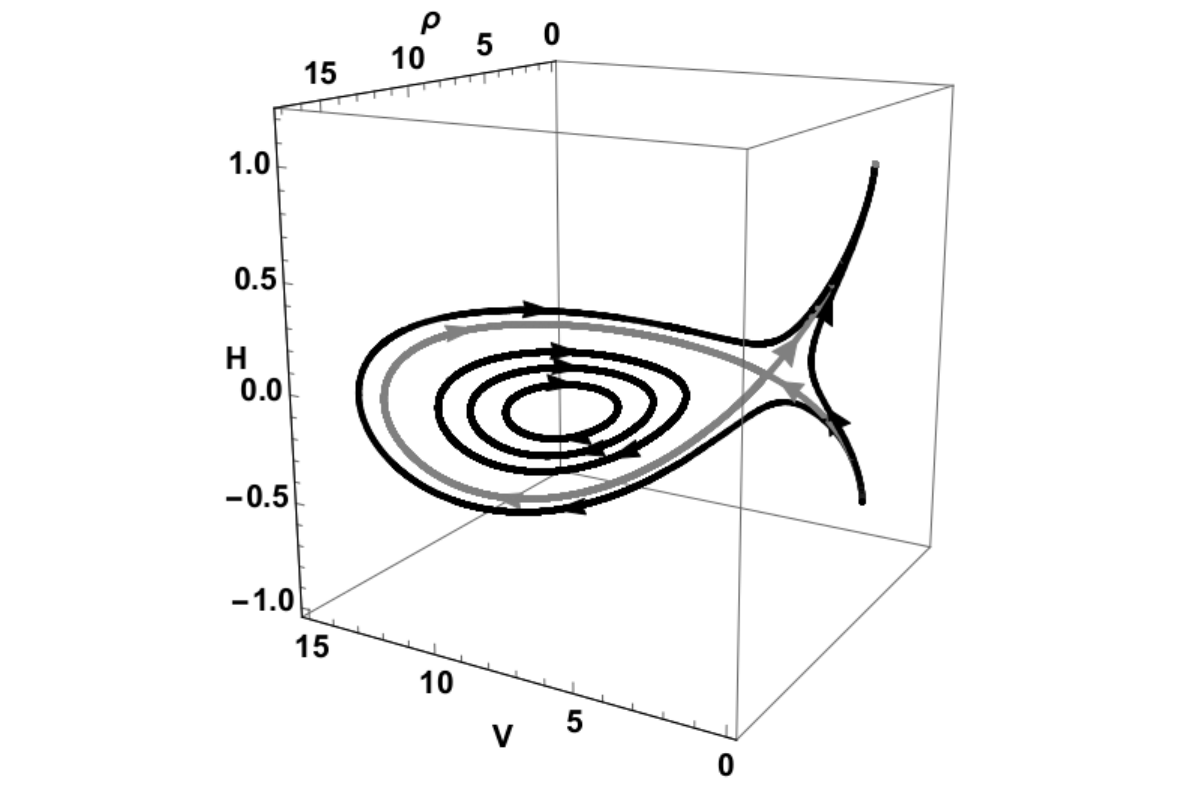} 
\caption{The three-dimensional phase space $(\rho, V, H)$ with $k>0$ and non-linear coupling parameter $\chi=0.085$. The trajectories shown correspond to those shown in Figure~\ref{fig10} with first integral ${\mathcal K}_\Sigma=0$.}
\label{fig10b}
\end{figure}
\begin{figure}
\includegraphics[width=7.3cm,height=5cm]{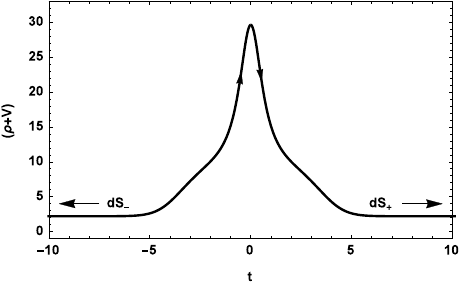}
\includegraphics[width=7.3cm,height=5cm]{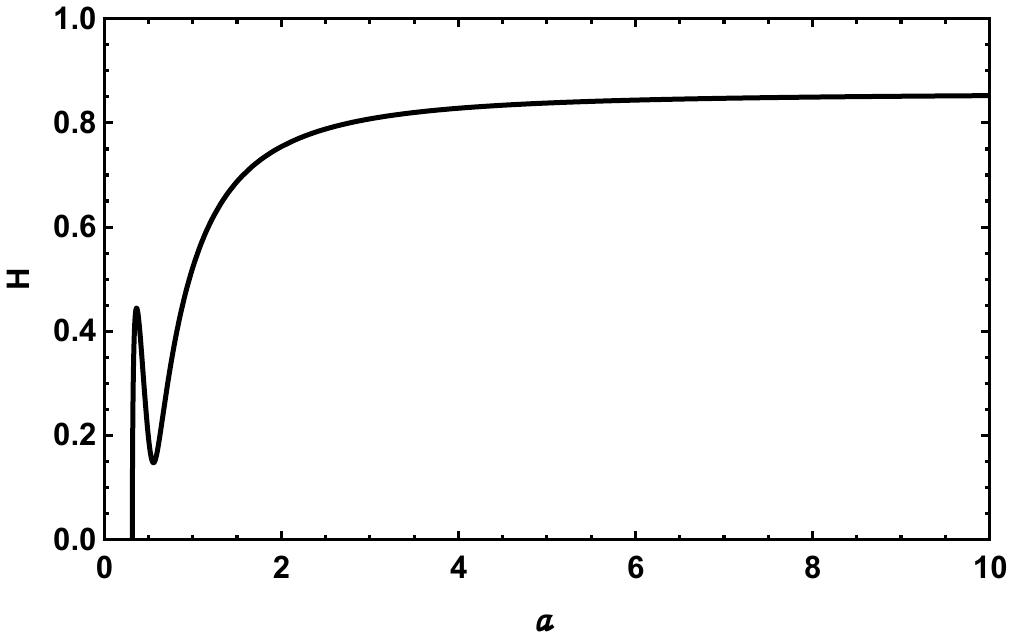}
\caption{The behaviour of $\rho+V$ as a function of $t$ (top panel) for a particular orbit in
region II of Fig. 8.
Here we see that the total energy density, $\rho+V$, is always positive throughout the bounce. On the bottom panel we show the behaviour of $H$ as a function of $a$ for the same orbit but with initial conditions
$a_\ast\simeq 0.31$, $V_\ast=13.4$ and $H_\ast=0$. Here we see that $H\rightarrow H_{\Lambda+}\simeq 0.85$ as $a\rightarrow +\infty$.}
\label{fig11}
\end{figure}

In order to examine the local structure of the phase space in a neighbourhood of Einstein fixed points,
we note that, for $w=0$, Eq.~(\ref{ev}) reduces to
\begin{eqnarray}
\label{evn}
\tilde {\gamma}_{\pm}=\kappa
\sqrt{\frac{V_E\left[(1+3\chi)V_\Lambda-3\chi V_E\right]}{V_\Lambda}}.
\end{eqnarray}
For $V_\Lambda >0$,
it is easy to show that $\tilde {\gamma}_\pm$ are pure imaginary for $V_E>V_{Emax}$.
That is, the fixed point obtained is a center.
For $V_\Lambda<V_E<V_{Emax}$ on the other hand, $\tilde {\gamma}_\pm$ is a pair of real eigenvalues. In this case, the fixed point
obtained is a saddle. Finally, in the limit $V_E\rightarrow V_{Emax}$ the two fixed points -- the centre and saddle -- pinch into
a fixed point with null eigenvalues which implies a bifurcation in the stability.

We shall restrict our analysis to the domain $0<\chi<\chi_{max}$ so that two fixed points (a center and a saddle) are present. As pointed out above, this is the configuration of interest in the sense that nonsingular models may be obtained. In fact, in Fig. \ref{fig10} (top panel) we show the phase portrait for several initial conditions in the case of $\chi=0.085$. From the saddle fixed point $P_2$ emerges a separatrix (grey curve)
dividing the phase space in three  distinct regions: region I, in a neighbourhood of the center
fixed point $P_1$, where perpetually cycling orbits  describe universe models going from contraction to expansion through a bounce, then re-contracting through a turn-around; regions II and III, both with  one bounce orbits. 
The corresponding orbits on the sector $(a, H)$ may be obtained as long as one increases the number of degrees
of freedom:
\begin{eqnarray}
\label{fds}
\dot{a}&=&H a,\\
\nonumber
\dot{V}&=&-3H\Big(1-\frac{V}{V_\Lambda}\Big)\\
&\times&\Big\{\chi V-\frac{V_\Lambda}{2}\ln\Big[\Big(1-\frac{V}{V_\Lambda}\Big)^2\Big]\Big\},\\
\dot{H}&=&-\frac{3H^2}{2}-\frac{k}{2a^2}+\frac{\kappa^2 V}{2}.
\end{eqnarray}
In this case the above dynamical system must be 
subjected to the Friedmann constraint (\ref{eqm9})
\begin{eqnarray}
H^2+\frac{k}{a^2}=\frac{\kappa^2 V_\Lambda}{6\chi}\ln\Big[\Big(1-\frac{V}{V_\Lambda}\Big)^2\Big].
\end{eqnarray}
It is also worth noting that Friedmann equation (\ref{eqm9}) 
evaluated at the Einstein fixed points gives the corresponding values of the scale factor $a_{E}$:
\begin{eqnarray}
\label{vf}
V_E=\frac{k(1+3w)}{\kappa^2(1+w)a^2_E}.
\end{eqnarray}
It can be easily seen from (\ref{vf}) that the corresponding Einstein fixed points
$P_1$ and $P_2$ can only be obtained as long as $k>0$.
On the bottom panel of Figure~\ref{fig10} we display the orbits in the $(a, H)$ sector assuming
the same initial conditions from the top panel. 
In Figure~\ref{fig10b} we show orbits in the full phase space $(\rho, V, H)$ corresponding to those of Figure~\ref{fig10}.

In Figure~\ref{fig11} (top panel) we show the behaviour of  $\rho+V$ as a function of $t$ for a particular orbit in
region II of Figure~\ref{fig10}.
Here we see that the total energy density $(\rho+V)$ is always positive
so that the weak energy condition is satisfied.
In addition, given the existence of the vacuum sub-manifold ($\rho=0$), if $\rho$ is initially positive then it remains positive.
On the bottom panel of Figure~\ref{fig11} we show the behaviour of $H$ as a function of $a$ for the same orbit. 
Here we see that the model approaches the de Sitter attractor as $a\rightarrow +\infty$, as expected.
That is, $H\rightarrow H_{\Lambda+}\simeq 0.85$ as $a\rightarrow +\infty$.

\section{Final Remarks}

In this paper we have considered FLRW cosmological models in which the vacuum energy interacts with a perfect fluid. 
We have considered both linear and non-linear couplings leading to an energy transfer between the two components. In particular we have investigated the existence of nonsingular solutions.

In our first example, the energy transfer is given by
a linear combination of the vacuum and fluid energy densities. In this linear case we integrate the coupled conservation equations obtaining the general solution for the energy densities of the matter and vacuum components. Substituting these into the Friedmann equation leads to a nonsingular evolution of the universe for some regions in parameter space. 
This can occur even for spatially flat or hyperbolic cosmologies with pressureless matter, for example, which is perhaps unexpected since the non-interacting vacuum plus matter, with equation of state $w\geq0$, would not exhibit non-singular behaviour. However the interaction can give rise to a term in the vacuum energy which, even if $V\to V_\Lambda>0$ at late times, for a sufficiently large coupling parameter, $\xi>1+w$, can act like a negative energy density with sufficiently stiff equation of state to produce an infinite barrier in the potential, $U(a)$ in Eq.~\eqref{eqa13}, at small values of the scale factor and thus generate a bounce.
It is interesting to point out that if the interaction term is sufficiently stiff,  the barrier and the bounce would persist in the presence of an additional non-interacting  component. 
\revised{For example, including radiation, $\rho_R\propto a^{-4}$, it is clear from \eqref{eqa05} that the interaction between pressureless matter and the vacuum can lead  to a bounce for $\xi>4/3$.}

\revised{Such a large dimensionless coupling between the matter and vacuum at early times can violate observational bounds on the allowed coupling at late times in simple linear interaction models. In Ref.~\cite{Martinelli:2019dau}, for example, a linear interaction with $\sigma=0$ and $V_\Lambda=0$ in Eq.~(\ref{eqm111}) is studied giving bounds on $|\xi|<0.06$ at 95\%~c.l. One might expect this bound to be relaxed in the more general cases with non-zero $\sigma$ and/or $V_\Lambda$, but it is important to realise that this observational bound comes from data at low redshifts and hence low energies. Studies have shown that in the same model there is no bound from observational data on the value of the coupling at high redshifts, $z>2.5$~\cite{Salvatelli:2014zta,Martinelli:2019dau,Hogg:2020rdp}. This motivates us to look at interactions beyond the simplest linear case, in which case the effective coupling may differ between high and low energies.}

In our second example, the energy transfer includes
the product of the vacuum and fluid energy densities.
In this nonlinear case, we can obtain a first integral of the conservation equations, enabling us to investigate the existence of 
nonsingular cosmologies. Again, for some range of parameter values we find nonsingular solutions.

In both cases conditions for the existence of a bounce give the same topology in the $(a,H)$ phase space, 
leading to a qualitatively similar behaviour as illustrated by Figures~\ref{fig2} and \ref{fig10}. 
The phase space shows the existence of nonsingular orbits with two accelerated phases, 
separated by a smooth transition corresponding to a decelerated expansion. 
Although we have focussed on the example of a single pressureless matter fluid interacting with the vacuum, we expect to see similar non-singular behaviour for sufficiently strong coupling in the presence of other components, notably radiation which must dominate the cosmic expansion at high energies, e.g., during primordial nucleosynthesis.

Previous works have explored the observational constraints on interacting vacuum cosmologies for particular interaction models. In general couplings are constrained to be small in the late-time universe where the $\Lambda$CDM model provides a good fit to data
\cite{Quercellini:2008vh,Wang:2013qy,Wang:2015wga,Sola:2016sbt,DiValentino:2017iww,Martinelli:2019dau,Hogg:2020rdp,SolaPeracaula:2021gxi}.
However as far as we are aware there are no current bounds on the specific form of nonlinear coupling studied here, where the interaction is naturally suppressed at late times and we recover an asymptotic accelerated de Sitter expansion \cite{Bruni:2001pc}.

\revised{We note that nonsingular solutions can also be found in 
the framework of the so-called Running Vacuum Models \cite{Lima:2013dmf,Lima:2014hia,SolaPeracaula:2019kfm}. In these models
the vacuum component may be realized as the sum of even powers of the Hubble expansion rate, 
following quantum field theory arguments in curved spacetime \cite{Moreno-Pulido:2020anb}.
For the particular case in which the nonsingular term is absent in the vacuum energy density,
the RVM and the models explored in section~III are compared against observational data in Ref.~\cite{SolaPeracaula:2017esw}.}

In future work we intend to examine the phenomenology
of the bounce including the spectrum of scalar perturbations that could be generated approaching the bounce in models that remain consistent with cosmological observations at late times. In this case, modifications to the equations for cosmological perturbations might furnish interesting predictions about the growth of structure formation~\cite{Borges:2017jvi}.

Finally, an extension of the interacting models considered here  which deserves further examination is the case of 
a general anisotropic Bianchi IX cosmology with $3$ scale factors.
In this case Einstein's equations reduce to a dynamical system with more degrees of freedom
furnishing a richer dynamics. Taking into account the
interacting terms, an Einstein fixed point (a saddle-center-center in the six-dimensional phase-space) may be obtained
in the case of nonsingular configurations. 
From such fixed points one might obtain stable and unstable 4-dim cylinders in which
oscillatory motions about the separatrix take place towards the bounce so that the homoclinic transversal intersection
furnishes an invariant chaos signature for the models.
As shown in \cite{Maier:2015kma, Maier:2017dtb}, this behavior defines a {\it chaotic saddle} indicating that the intersection points of the cylinders have the nature of a Cantor set. We also intend to examine the possibly-oscillatory approach to the bounce and analogous features present in the BKL conjecture in general relativity~\cite{Belinsky:1970ew, Khalatnikov:1969eg, Belinsky:1982pk}.  In the context of Bianchi IX models, which may be regarded as providing the general spatially-averaged description of the Universe, the challenge \cite{Bozza:2009jx} is to suppress anisotropy enough during the pre-bounce collapsing phase, in order to generate a viable post-bounce cosmology. Such suppressing mechanisms have been shown to exist \cite{Ganguly:2019llh}; in future we intend to investigate if   the type of interactions considered here can naturally provide, in the context of anisotropic Bianchi IX models, both the bounce and the mechanisms to make it sufficiently isotropic.     

~
\medskip

\begin{acknowledgments}
MB and DW are supported by UK STFC Grant No. ST/S000550/1. RM is supported by 
Coordenação
de Aperfeiçoamento de Pessoal de Nível Superior-Brasil (CAPES)-Código de Financiamento 001.
\end{acknowledgments}


\begin{thebibliography}{99}

\bibitem{weinbergtb} S. Weinberg, {\it Cosmology} (Oxford University Press, 2008).

\bibitem{maartenstb} George F. R. Ellis, Roy Maartens and Malcolm A. H. Maccallum, {\it Relativistic Cosmology}
(Cambridge University Press, 2012).

\bibitem{Jones:2017xsc}
B.~J.~T.~Jones,
{\it Precision Cosmology} (Cambridge University Press, 2017).

\bibitem{Gasperini:1992em}
M.~Gasperini and G.~Veneziano,
Astropart. Phys. \textbf{1}, 317-339 (1993)
[arXiv:hep-th/9211021 [hep-th]].

\bibitem{Khoury:2001wf}
J.~Khoury, B.~A.~Ovrut, P.~J.~Steinhardt and N.~Turok,
Phys. Rev. D \textbf{64}, 123522 (2001)
[arXiv:hep-th/0103239 [hep-th]].

\bibitem{Kallosh:2001ai}
R.~Kallosh, L.~Kofman and A.~D.~Linde,
Phys. Rev. D \textbf{64}, 123523 (2001)
[arXiv:hep-th/0104073 [hep-th]].

\bibitem{Steinhardt:2001st}
P.~J.~Steinhardt and N.~Turok,
Phys. Rev. D \textbf{65}, 126003 (2002)
[arXiv:hep-th/0111098 [hep-th]].

\bibitem{Wands:2008tv}
D.~Wands,
Adv. Sci. Lett. \textbf{2}, 194-204 (2009)
[arXiv:0809.4556 [astro-ph]].

\bibitem{Maier:2013hr}
R.~Maier, N.~Pinto-Neto and I.~D.~Soares,
Phys. Rev. D \textbf{87}, no.4, 043528 (2013)
[arXiv:1301.5250 [gr-qc]].

\bibitem{Maier:2011yy}
R.~Maier, S.~Pereira, N.~Pinto-Neto and B.~B.~Siffert,
Phys. Rev. D \textbf{85}, 023508 (2012)
[arXiv:1111.0946 [astro-ph.CO]].

\bibitem{SupernovaSearchTeam:1998fmf}
A.~G.~Riess \textit{et al.} [Supernova Search Team],
Astron. J. \textbf{116}, 1009-1038 (1998)
[arXiv:astro-ph/9805201 [astro-ph]].

\bibitem{SupernovaCosmologyProject:1998vns}
S.~Perlmutter \textit{et al.} [Supernova Cosmology Project],
Astrophys. J. \textbf{517}, 565-586 (1999)
[arXiv:astro-ph/9812133 [astro-ph]].

\bibitem{Rubin:2008wq}
D.~Rubin, E.~V.~Linder, M.~Kowalski, G.~Aldering, R.~Amanullah, K.~Barbary, N.~V.~Connolly, K.~S.~Dawson, L.~Faccioli and V.~Fadeyev, \textit{et al.}
Astrophys. J. \textbf{695}, 391-403 (2009)
[arXiv:0807.1108 [astro-ph]].

\bibitem{Hicken:2009dk}
M.~Hicken, W.~M.~Wood-Vasey, S.~Blondin, P.~Challis, S.~Jha, P.~L.~Kelly, A.~Rest and R.~P.~Kirshner,
Astrophys. J. \textbf{700}, 1097-1140 (2009)
[arXiv:0901.4804 [astro-ph.CO]].

\bibitem{BOSS:2016wmc}
S.~Alam \textit{et al.} [BOSS],
Mon. Not. Roy. Astron. Soc. \textbf{470}, no.3, 2617-2652 (2017)
[arXiv:1607.03155 [astro-ph.CO]].

\bibitem{Planck:2018vyg}
N.~Aghanim \textit{et al.} [Planck],
Astron. Astrophys. \textbf{641}, A6 (2020)
[erratum: Astron. Astrophys. \textbf{652}, C4 (2021)]
[arXiv:1807.06209 [astro-ph.CO]].

\bibitem{Hawking:1973uf}
S.~W.~Hawking and G.~F.~R.~Ellis,
{\it The Large Scale Structure of Space-Time} 
(Cambridge University Press, 1973).

\bibitem{Wald:1984rg}
R.~M.~Wald,
{\it General Relativity} (University of Chicago Press, 1984).

\bibitem{Visser:1997qk}
M.~Visser,
Science \textbf{276}, 88-90 (1997)
[arXiv:1501.01619 [gr-qc]].

\bibitem{Weinberg:1988cp}
S.~Weinberg,
Rev. Mod. Phys. \textbf{61}, 1-23 (1989).

\bibitem{amendola} Luca Amendola and Shinji Tsujikawa, 
{\it Dark Energy: Theory and Observations}, (Cambridge University Press, 2010).

\bibitem{Carneiro:2009et}
S.~Carneiro and R.~Tavakol,
Phys. Rev. D \textbf{80}, 043528 (2009)
[arXiv:0907.4795 [astro-ph.CO]].

\bibitem{Chimento:2014tua}
L.~P.~Chimento and S.~Carneiro,
AIP Conf. Proc. \textbf{1647}, no.1, 10-12 (2015)
[arXiv:1402.2311 [astro-ph.CO]].

\bibitem{Ananda:2005xp}
K.~N.~Ananda and M.~Bruni,
Phys. Rev. D \textbf{74}, 023523 (2006).
[arXiv:astro-ph/0512224 [astro-ph]].

\bibitem{Ananda:2006gf}
K.~N.~Ananda and M.~Bruni,
Phys. Rev. D \textbf{74}, 023524 (2006)
[arXiv:gr-qc/0603131 [gr-qc]].


\bibitem{Ganguly:2019llh}
C.~Ganguly and M.~Bruni,
Phys. Rev. Lett. \textbf{123}, no.20, 201301 (2019)
[arXiv:1902.06356 [gr-qc]].



\bibitem{Bertolami:1986bg}
O.~Bertolami,
Nuovo Cim. B \textbf{93}, 36-42 (1986)

\bibitem{Alcaniz:2012mh}
J.~S.~Alcaniz, H.~A.~Borges, S.~Carneiro, J.~C.~Fabris, C.~Pigozzo and W.~Zimdahl,
Phys. Lett. B \textbf{716}, 165-170 (2012)
[arXiv:1201.5919 [astro-ph.CO]].

\bibitem{Freese:1986dd}
K.~Freese, F.~C.~Adams, J.~A.~Frieman and E.~Mottola,
Nucl. Phys. B \textbf{287}, 797-814 (1987).

\bibitem{Carvalho:1991ut}
J.~C.~Carvalho, J.~A.~S.~Lima and I.~Waga,
Phys. Rev. D \textbf{46}, 2404-2407 (1992).

\bibitem{Shapiro:2000dz}
I.~L.~Shapiro and J.~Sola,
JHEP \textbf{02}, 006 (2002)
[arXiv:hep-th/0012227 [hep-th]].

\bibitem{Wands:2012vg}
D.~Wands, J.~De-Santiago and Y.~Wang,
Class. Quant. Grav. \textbf{29}, 145017 (2012)
[arXiv:1203.6776 [astro-ph.CO]].


\bibitem{Chen:1990jw}
\revised{W.~Chen and Y.~S.~Wu,
Phys. Rev. D \textbf{41}, 695-698 (1990)
[erratum: Phys. Rev. D \textbf{45}, 4728 (1992)]
doi:10.1103/PhysRevD.41.695
}


\bibitem{Lima:2013dmf}
\revised{J.~A.~S.~Lima, S.~Basilakos and J.~Sola,
Mon. Not. Roy. Astron. Soc. \textbf{431}, 923-929 (2013)
[arXiv:1209.2802 [gr-qc]].
}

\bibitem{Moreno-Pulido:2020anb}
\revised{C.~Moreno-Pulido and J.~Sola,
Eur. Phys. J. C \textbf{80}, no.8, 692 (2020)
[arXiv:2005.03164 [gr-qc]].
}

\bibitem{Maier:2020bgm}
R.~Maier,
Int. J. Mod. Phys. D \textbf{29}, no.14, 2043023 (2020)
[arXiv:2005.09576 [gr-qc]].

\bibitem{Salvatelli:2014zta}
V.~Salvatelli, N.~Said, M.~Bruni, A.~Melchiorri and D.~Wands,
Phys. Rev. Lett. \textbf{113}, no.18, 181301 (2014)
[arXiv:1406.7297 [astro-ph.CO]].

\bibitem{Benetti:2021div}
M.~Benetti, H.~Borges, C.~Pigozzo, S.~Carneiro and J.~Alcaniz,
JCAP \textbf{08}, 014 (2021)
[arXiv:2102.10123 [astro-ph.CO]].

\bibitem{Wang:2015wga}
Y.~Wang, G.~B.~Zhao, D.~Wands, L.~Pogosian and R.~G.~Crittenden,
Phys. Rev. D \textbf{92}, 103005 (2015)
[arXiv:1505.01373 [astro-ph.CO]].

\bibitem{Zhao:2017cud}
G.~B.~Zhao, M.~Raveri, L.~Pogosian, Y.~Wang, R.~G.~Crittenden, W.~J.~Handley, W.~J.~Percival, F.~Beutler, J.~Brinkmann and C.~H.~Chuang, \textit{et al.}
Nature Astron. \textbf{1}, no.9, 627-632 (2017)
[arXiv:1701.08165 [astro-ph.CO]].

\bibitem{Sola:2016sbt}
J.~Sol\`a, A.~G\'omez-Valent and J.~de Cruz P\'erez,
Int. J. Mod. Phys. A \textbf{32}, no.19-20, 1730014 (2017)
[arXiv:1709.07451 [astro-ph.CO]].

\bibitem{DiValentino:2017iww}
E.~Di Valentino, A.~Melchiorri and O.~Mena,
Phys. Rev. D \textbf{96}, no.4, 043503 (2017)
[arXiv:1704.08342 [astro-ph.CO]].

\bibitem{Kumar:2017dnp}
S.~Kumar and R.~C.~Nunes,
Phys. Rev. D \textbf{96}, no.10, 103511 (2017)
[arXiv:1702.02143 [astro-ph.CO]].

\bibitem{Wang:2013qy}
Y.~Wang, D.~Wands, L.~Xu, J.~De-Santiago and A.~Hojjati,
Phys. Rev. D \textbf{87}, no.8, 083503 (2013)
[arXiv:1301.5315 [astro-ph.CO]].

\bibitem{Martinelli:2019dau}
M.~Martinelli, N.~B.~Hogg, S.~Peirone, M.~Bruni and D.~Wands,
Mon. Not. Roy. Astron. Soc. \textbf{488}, no.3, 3423-3438 (2019)
[arXiv:1902.10694 [astro-ph.CO]].

\bibitem{Hogg:2020rdp}
N.~B.~Hogg, M.~Bruni, R.~Crittenden, M.~Martinelli and S.~Peirone,
Phys. Dark Univ. \textbf{29}, 100583 (2020)
[arXiv:2002.10449 [astro-ph.CO]].

\bibitem{DiValentino:2021izs}
E.~Di Valentino, O.~Mena, S.~Pan, L.~Visinelli, W.~Yang, A.~Melchiorri, D.~F.~Mota, A.~G.~Riess and J.~Silk,
Class. Quant. Grav. \textbf{38}, no.15, 153001 (2021)
[arXiv:2103.01183 [astro-ph.CO]].



\bibitem{Hogg:2021yiz}
N.~B.~Hogg and M.~Bruni,
[arXiv:2109.08676 [astro-ph.CO]].


\bibitem{SolaPeracaula:2021gxi}
J.~Sol\`a Peracaula, A.~G\'omez-Valent, J.~de Cruz Perez and C.~Moreno-Pulido,
EPL \textbf{134}, no.1, 19001 (2021)
[arXiv:2102.12758 [astro-ph.CO]].



\bibitem{Kaeonikhom:2020fqs}
C.~Kaeonikhom, P.~Rangdee, H.~Assadullahi, B.~Gumjudpai, J.~A.~Schewtschenko and D.~Wands,
Phys. Rev. D \textbf{102}, 123519 (2020)
[arXiv:2007.12181 [astro-ph.CO]].


\bibitem{Quercellini:2008vh}
C.~Quercellini, M.~Bruni, A.~Balbi and D.~Pietrobon,
Phys. Rev. D \textbf{78}, 063527 (2008)
[arXiv:0803.1976 [astro-ph]].



\bibitem{AP}
D. K. Arrowsmith and C. M. Place, {\it Dynamical systems: differential equations, maps and chaotic behaviour} (Chapman and Hall, London, 1992).
        

\bibitem{Parisi:2007kv}
L.~Parisi, M.~Bruni, R.~Maartens and K.~Vandersloot,
Class. Quant. Grav. \textbf{24}, 6243-6254 (2007)
[arXiv:0706.4431 [gr-qc]].


\bibitem{DiValentino:2019qzk}
E.~Di Valentino, A.~Melchiorri and J.~Silk,
Nature Astron. \textbf{4}, no.2, 196-203 (2019)
[arXiv:1911.02087 [astro-ph.CO]].

\bibitem{Handley:2019tkm}
W.~Handley,
Phys. Rev. D \textbf{103}, no.4, L041301 (2021)
[arXiv:1908.09139 [astro-ph.CO]].

\bibitem{Peter:2006hx}
P.~Peter, E.~J.~C.~Pinho and N.~Pinto-Neto,
Phys. Rev. D \textbf{75}, 023516 (2007)
[arXiv:hep-th/0610205 [hep-th]].

\bibitem{Pinto-Neto:2009kyd}
N.~Pinto-Neto,
Phys. Rev. D \textbf{79}, 083514 (2009)
[arXiv:0904.4454 [gr-qc]].



\bibitem{Lesgourgues:2006nd}
J.~Lesgourgues and S.~Pastor,
Phys. Rept. \textbf{429}, 307-379 (2006)
[arXiv:astro-ph/0603494 [astro-ph]].


\bibitem{Bruni:2001pc}
M.~Bruni, F.~C.~Mena and R.~K.~Tavakol,
Class. Quant. Grav. \textbf{19}, L23-L29 (2002)
[arXiv:gr-qc/0107069 [gr-qc]].



{\color{black}
\bibitem{Lima:2014hia}
J.~A.~S.~Lima, S.~Basilakos and J.~Sol\`a,
Gen. Rel. Grav. \textbf{47}, 40 (2015)
[arXiv:1412.5196 [gr-qc]].

\bibitem{SolaPeracaula:2019kfm}
J.~Sol\`a Peracaula and H.~Yu,
Gen. Rel. Grav. \textbf{52}, no.2, 17 (2020)
[arXiv:1910.01638 [gr-qc]].

\bibitem{SolaPeracaula:2017esw}
J.~Sol\`a Peracaula, J.~de Cruz P\'erez and A.~Gomez-Valent,
Mon. Not. Roy. Astron. Soc. \textbf{478}, no.4, 4357-4373 (2018)
[arXiv:1703.08218 [astro-ph.CO]].
}


\bibitem{Borges:2017jvi}
H.~A.~Borges and D.~Wands,
Phys. Rev. D \textbf{101}, no.10, 103519 (2020)
[arXiv:1709.08933 [astro-ph.CO]].



\bibitem{Maier:2015kma}
R.~Maier, I.~D.~Soares and E.~V.~Tonini,
Class. Quant. Grav. \textbf{32}, no.23, 235001 (2015)
[arXiv:1505.06189 [gr-qc]].

\bibitem{Maier:2017dtb}
R.~Maier and I.~D.~Soares,
Phys. Rev. D \textbf{96}, no.10, 103532 (2017)
[arXiv:1710.06911 [gr-qc]].

\bibitem{Belinsky:1970ew}
V.~A.~Belinsky, I.~M.~Khalatnikov and E.~M.~Lifshitz,
Adv. Phys. \textbf{19}, 525-573 (1970).

\bibitem{Khalatnikov:1969eg}
I.~M.~Khalatnikov and E.~M.~Lifshitz,
Phys. Rev. Lett. \textbf{24}, 76-79 (1970).

\bibitem{Belinsky:1982pk}
V.~a.~Belinsky, I.~m.~Khalatnikov and E.~m.~Lifshitz,
Adv. Phys. \textbf{31}, 639-667 (1982).





\bibitem{Bozza:2009jx}
V.~Bozza and M.~Bruni,
JCAP \textbf{10}, 014 (2009)
[arXiv:0909.5611 [hep-th]].


\end{thebibliography}
\end{document}